\documentclass[11pt]{article}

\usepackage{graphicx}
\usepackage{epstopdf}
\usepackage{bm}
\usepackage{color}

\newcommand{\be}[1]{\begin{equation}\label{#1}}
\newcommand{\ee}{\end{equation}}

\usepackage{mathtools}

\DeclareMathOperator{\re}{Re}

\def \a {\alpha}
\def \b {\beta}
\def \g {\gamma}
\def \gat {\tilde{\gamma}}
\def \d {\delta}
\def \e {\epsilon}
\def \ve {\varepsilon}
\def \vet {\tilde{\varepsilon}}

\def \m {\mu}
\def \n {\nu}

\def \l {\lambda}

\def \detg {\sqrt{-g}}

\def \E {{\boldmath $E$}}
\def \B {{\boldmath $B$}}
\def \D {{\boldmath $D$}}
\def \H {{\boldmath $H$}}

\def \Em {\text{\E}}
\def \Bm {\text{\B}}
\def \Dm {\text{\D}}
\def \Hm {\text{\H}}

\begin{document}

\begin{titlepage}
\vfill

\begin{center}
\baselineskip=16pt
{\Large\bf 
A differential geometry approach to asymmetric\\transmission of light
}
\vskip 0.5cm

{\large{Felipe A. Asenjo$^1$, Cristi\'an Erices$^2$, Andr\'es Gomberoff$^{1,2}$, Sergio A. Hojman$^{3,4}$, and Alejandra Montecinos$^{5}$}}

\textit{$^1$Facultad de Ingenier\'ia y Ciencias, Universidad Adolfo Ib\'a\~nez, Av.~Diagonal las Torres 2640, Pe\~nalol\'en, Santiago, Chile\\
$^2$Centro de Estudios Cient\'{\i}ficos (CECs), Av. Arturo Prat 514, Valdivia, Chile\\
$^3$Facultad de Artes Liberales, Universidad Adolfo Ib\'a\~nez, Av.~Diagonal las Torres 2640, Pe\~nalol\'en, Santiago, Chile\\
$^4$Departamento de F\'isica, Facultad de Ciencias, Universidad de Chile, Las Palmeras 3425, \~Nu\~noa, Chile\\
$^5$Universidad Andres Bello,
Departamento de Ciencias F\'{\i}sicas,
Av. Rep\'ublica 220, Santiago, Chile
}
\vspace{6pt}
\end{center}
\vskip 0.2in
\par
\begin{center}
{\bf Abstract}
\end{center}


\begin{quote}
In the last ten years, the technology of differential geometry, ubiquitous in gravitational physics, has found its place in the field of optics. It has been successfully used in the design of optical metamaterials, through a technique now known as ``transformation optics''. This method, however, only applies for the particular class of metamaterials known as impedance matched, that is, materials  whose electric permittivity is equal to their magnetic permeability. In that case, the material may be described by a spacetime metric. In the present work we will introduce a generalization of the geometric methods of transformation optics to situations in which the material is not impedance matched. In such situation, the material -or more precisely, its constitutive tensor- will not be described by a metric only. We bring in a second tensor, with the local symmetries of the Weyl tensor, the ``$W$-tensor". In the geometric optics approximation we show how the properties of the $W$-tensor are related to the asymmetric transmission of the material. We apply this feature into the design of a particularly interesting set of asymmetric materials. These materials are birefringent when light rays approach the material in a given direction, but behave just like vacuum when the rays have the opposite direction with the appropriate polarization (or, in some cases, independently of the polarization).
\vfill
\vskip 2.mm
\end{quote}
\end{titlepage}

\tableofcontents

\newpage 

\section{Introduction}

The last decade has witnessed the development of several novel and powerful methods for designing devices that may control electromagnetic radiation in many useful ways.  Theoretically, the goal is to find a prescription for the construction of a material medium, given by its constitutive tensor, so that light will behave with some desired, given properties. One then may use it to design invisibility devices, perfect lenses, or optical black holes \cite{ulf4,gbur,uhlmann1,rosquist,alu,ulf2,pendry3,ulf3,pendry1,milton,ulf1,pendry2,chen}.
A particularly successful technique, is known as  ``transformation optics''\cite{pendry1, ulf1, ulf3}. It borrows one of the mathematical tools mostly used in  General Relativity, namely, differential geometry. It has been specially useful in the design of invisibility devices \cite{chenchan}. However, transformation optics can only deal with a particular family of materials, namely, the ones satisfying the ``impedance matching" condition,
\begin{equation}\label{im}
\mu_{ij}=\epsilon_{ij} ,
\end{equation}
where $\mu_{ij}$, $\epsilon_{ij}$,  are the magnetic permeability and electric permittivity matrices, respectively.

  The impedance matching condition restricts the class of materials that may be discussed by transformation optics. The purpose of this note is to extend the geometric treatment of optical metamaterials to cases where (\ref{im}) is not fulfilled. This is achieved by the introduction of a new tensor, the $W$-tensor, with the local properties of the Weyl tensor of General Relativity. However, this new object is not the Weyl tensor, in the sense that it is  not related to a metric, but it can be chosen freely. Its degrees of freedom are the ones missing in a metric treatment of an optical material (in the way transformation optics does) contributing to its constitutive tensor.

 Within the validity domain of geometric optics we will analyze the generic case of linearly polarized electromagnetic waves propagating in a homogeneous material taking advantage of the algebraic properties of the $W$-tensor.

 For a homogeneous and time independent medium described by transformation optics, one may always perform a Lorentz transformations such that the components $g_{i0}$ of the metric describing it vanish. In such a reference frame the geometry is static, and therefore the geodesics are time invertible and the light transmission is symmetric. This is no longer true for generic homogeneous and static media, and we will analyze them paying particular attention to this feature.
 
The technique may allow, for example, the design of a material sensitive to the direction of light propagation, which may be useful in the development of new classes of optical diodes or other devices showing asymmetric transmission of light (see, for instance, \cite{PhysRevLett.108.213905,Mutlu:11} for recent experimental developments on these materials)

The paper is organized as follows. In Sec. \ref{MaxCurved} we review the covariant formulation of electromagnetism in a medium, which is relevant for what follows. Then, in Sec. \ref{noveld} we show how ``transformation optics'' may be generalized with the introduction of the $W$-tensor, so that the constitutive tensor of Maxwell equations includes all its possible degrees of freedom. In Sec. \ref{Lightrays} we write down Maxwell equations in the geometrical optics limit, and show that light propagates as if it were in vacuum in some particular directions which may be read from the algebraic properties of the $W$-tensor. Then, in Sec. \ref{PetrovNmaterialSe} we study the transmission of light in the reverse direction. Finally, in Sec. \ref{discusion} we discuss our findings.

\section{Covariant formulation of electromagnetism. A review}
\label{MaxCurved}

In order to set up the notation,  we will start with a short review of the  covariant formulation of Maxwell's theory in a linear media and transformation optics. The action principle of the electromagnetic field is
\begin{equation}\label{action}
I=-\frac{1}{4} \int d^4 x \ \chi^{\alpha\beta\gamma\delta} F_{\alpha\beta}F_{\gamma\delta},
\end{equation}
where $F_{\mu\nu}=\partial_\mu A_\nu - \partial_\nu A_\mu$ is the electromagnetic field,  $A_\mu$ is the electromagnetic potential, and $\partial_\mu$ is a partial derivative (greek indices running from 0 to 3). The constitutive tensor $\chi^{\alpha\beta\gamma\delta}$ encodes all the electromagnetic properties of the material, and from the action one sees that it must have the following symmetries,
\begin{equation}\label{symm}
\chi^{\alpha\beta\gamma\delta}=-\chi^{\beta\alpha\gamma\delta}=\chi^{\beta\alpha\delta\gamma},
\ \ \ \ \ \ \chi^{\beta\alpha\gamma\delta}=\chi^{\gamma\delta\beta\alpha}.
\end{equation}
These imply that $\chi^{\alpha\beta\gamma\delta}$ has 21 independent parameters, namely, the 6 components of the electric permittivity, the 6 of the magnetic permeability and the 9 of the magneto-electric coupling.

Varying  action \eqref{action} one obtains the source-free Maxwell equations in a a dielectric media,
\begin{align}\label{Meq}
F_{\m\n,\l}+F_{\n\l,\m}+F_{\l\m,\n}&=0, & {H^{\m\n}}_{,\n}=0,
\end{align}
where
\begin{equation}
H^{\m\n}=\frac{1}{2}\chi^{\mu\nu\alpha\beta}F_{\alpha\beta}
\end{equation}
is the electromagnetic displacement tensor, constructed from the electric displacement $\vec D$ and magnetic field $\vec H$ in the same way as $F^{\mu\nu}$ is built from the electric field $\vec E$ and the magnetic induction $\vec B$. The technique known as transformations optics stems from the fact that action (\ref{action}) describes an electromagnetic field on a background geometry prescribed by certain metric tensor $g_{\mu\nu}$ if one identifies
\begin{equation}\label{metricdecomp}
 \chi^{\mu\nu\alpha\beta}=\frac{1}{2}\sqrt{-g}\ (g^{\m\a}g^{\n\b} - g^{\m\b}g^{\n\a}).
\end{equation}
Therefore, if a particular constitutive tensor may be written as in \eqref{metricdecomp}, then we will be able to make use of the mathematical technology of differential geometry in order to study the electromagnetic properties of the system. There is, however, an important difference in the interpretation. Here, two metrics - not two geometries - represent two different materials, because we assume the background geometry to be always flat. The metric $g_{\mu\nu}$ is a background field entering the equations as for Maxwell equations on curved backgrounds, but it represents the properties of the material, not of the geometry. That is the reason why transformation optics is so useful. One may deform the coordinates in any way one likes, and the new metric will correspond to a new material.  One may readily compute the trajectories of light rays. They are just the light-like geodesics of the metric $g_{\mu\nu}$. One may, therefore, apply transformations of coordinates to transform a set of light rays, into a different one satisfying some required properties. The transformed constitutive tensor will describe the media needed for getting those rays. This has been used, for instance, in the design of invisibility devices \cite{pendry1} (for a review see \cite{chenchan}).

Of course, not every constitutive tensor may be written as in Eq. \eqref{metricdecomp}. The metric tensor has 10 independent parameters. Moreover, a conformal transformation of the metric in \eqref{metricdecomp} leaves the constitutive tensor invariant, therefore this is a family describing only 9 of the 21 parameters of a generic constitutive tensor. We will call a medium of this kind, a metric material. Such materials must satisfy the impedance matching condition \eqref{im}, and its electric-magnetic coupling is restricted from 9 to only 3 degrees of freedom.

\section{A novel decomposition}
\label{noveld}

Now we extend the geometric treatment of optical materials generalizing the form of the constitutive tensor. Doing so, a wider realm of dielectric media may be treated, relaxing the impedance matching condition.

A natural extension of \eqref{metricdecomp}, that includes all of its degrees of freedom is
 \begin{equation}\label{decomp}
 \chi^{\a\b\g\d}=\frac{1}{2}\ \sqrt{-g}\ \Phi (g^{\a\g}g^{\b\d} - g^{\a\d}g^{\b\g}) +  \frac{1}{2}\sqrt{-g}W^{\a\b\g\d}.
\end{equation}
 Here, $\Phi$ is a scalar field and $W^{\a\b\g\d}$ is a traceless tensor,
 \begin{equation}\label{trace}
 W^{\a\b\g\d}g_{\b\d}=0,
 \end{equation}
 with the same symmetries  of the constitutive tensor, given in \eqref{symm}. Note that this is quite reminiscent of the Ricci decomposition of the Riemann tensor. In fact, the Riemann tensor also satisfies \eqref{symm} and the tensor $W^{\a\b\g\d}$ is the analog of the Weyl tensor, being the traceless part of $\chi^{\a\b\g\d}$. However, it is {\it not} the Weyl tensor, because it has no relation with $g_{\mu\nu}$ whatsoever. It only shares its symmetries under index permutations and its tracelessness. No direct geometrical interpretation may be derived from it. There are two other important differences between \eqref{decomp} and the Ricci decomposition. First, the constitutive tensor has 21 parameters, one more than de Riemann tensor. This is because the latter also satisfies the Bianchi identity. In a dielectric material, this restriction is also demanded in some cases \cite{post}, but we are not going to require it here. In the decomposition \eqref{decomp}, the Bianchi identity would translate into the same restriction on $W^{\a\b\g\d}$, which we are not going to demand.   Also note that \eqref{decomp} seems to lack the ``semi-traceless" part of the Ricci decomposition. This is because the decomposition is not quite the same. Here, the metric is not a given, background field, but it takes 9 of the degrees of freedom of the constitutive tensor. These, along with the degree of freedom of the scalar field and the 11 of $W^{\a\b\g\d}$ give rise to the 21 parameters in $ \chi^{\a\b\g\d}$. The decomposition \eqref{decomp} was first studied in \cite{erices}. It was also used in the context of bimetric QED in \cite{Drummond:2016ukf}.

 Now, one may wonder if the parameters in \eqref{symm}  are, in fact, independent and if the relation is invertible. That is, if $\chi^{\a\b\g\d}$ is known, may we find a unique set $(\Phi,g_{\a\b} ,W^{\a\b\g\d})$ to decompose it as in \eqref{decomp}. To check it note that, from \eqref{decomp} and \eqref{trace},
 \begin{equation}\label{mash}
\chi^{\a\b\g\d}g_{\b\d}=\frac{3}{2}\sqrt{-g}\Phi g^{\a\g}.
\end{equation}
Here we have assumed that $g_{\a\b}$ is a metric, and therefore an invertible matrix. Contracting \eqref{mash} with the metric it is possible to find $\Phi$ in terms of $g_{\mu\nu}$,
 \begin{equation}\label{phi}
 \Phi=\frac{1}{6\sqrt{-g}}\chi^{\a\b\g\d}g_{\b\d}g_{\a\g}.
 \end{equation}
 Therefore, if we find a unique metric $g_{\a\b}$, then from the above equation we may find $\Phi$, and then directly from \eqref{decomp} we may solve for $W^{\a\b\g\d}$. Hence the problem is whether we can find a unique metric (up to scaling, as we already mentioned above). Plugging $\Phi$ from \eqref{phi} into Eq. \eqref{mash}, we obtain a system of equations for $g_{\a\b}$. Note that they are 9 independent equations, because tracing it will give rise to an identity. This is precisely what we need, because all of the equations are invariant under the scale of the metric, and therefore only 9 parameters of the metric are physical, as we expect. We may replace $\Phi$ in \eqref{phi} to write Eq. \eqref{mash} in the form
\begin{equation}\label{quadratic}
\left( \chi^{\a\b\g\d} \delta^\tau_\sigma - \frac{1}{4}\chi^{\a\b\tau\d} \delta^\gamma_\sigma\right) g_{\b\d} g_{\a\tau} = 0.
\end{equation}
This is a system of homogeneous second order equations for $g_{\a\b}$. We cannot be sure that non-zero solutions will exist for a given constitutive tensor. But this is not important for us. Note that in vacuum, or for any metric media the solution will be the particular metric. This means that at least in the vicinity of these known solutions we will be able to find a decomposition in the form of Eq. \eqref{decomp}.

In general, the electromagnetic tensor $F_{\a\b}$ contains the electric field $F_{0i}=E_{i}$, and the magnetic induction $F_{ij}=-\epsilon_{ijk}B^{k}$. The tensor density $H^{\m\n}$ contains the electric displacement $H^{0i}=-D^{i}$ and the magnetic field $H^{ij}=-\epsilon^{ijk}H_{k}$.
The  constitutive tensor \eqref{decomp} encodes the dielectric properties of the medium, i.e., its electric permittivity, magnetic permeability and bianisotropic properties in terms of the metric, the scalar field $\Phi$ and the $W$-tensor. The dielectric matrices may be obtained from the constitutive relations writing  $(\Dm,\Hm)$ in terms of $(\Em, \Bm)$. In that case we say that the constitutive relation is in the EB representation.  Similarly, if one writes ($\Dm,\Bm$) in terms of ($\Em$, $\Hm$), we refer to it as the EH representation. In the EB representation
\begin{subequations}\label{EBrep}
\begin{align}
\Dm&=\vet\Em+\gat\Bm\, ,\\
\Hm&=\m^{-1}\Bm-\gat^{t}\Em\ , 
\end{align}
\end{subequations}
where,
\begin{subequations}\label{ebmatrix}\begin{align}
\tilde{\varepsilon}^{ij}&=\detg \Phi (g^{0i}g^{0j}-g^{00}g^{ij})-\detg W^{0i0j}\, ,\\
\tilde{\gamma}^{i}_{\ j}&=\detg \Phi \e_{lkj}g^{0l}g^{ik}+\frac{1}{2}\detg \e_{lkj}W^{0ilk}\, ,\\
\mu^{-1}_{ij}&=\frac{1}{2} \Phi \detg \e_{ilk}\e_{jmn}g^{lm}g^{kn}+\frac{1}{4}\detg\e_{imn}\e_{jlk} W^{mnlk}\, .
\end{align}\end{subequations}
In the EH representation,
\begin{subequations}\label{EHrep}
\begin{align}
\Dm&=\ve \Em+\g \Hm \ , \\ 
\Bm&=\m \Hm+\g^{t}\Em \ ,
\end{align}
\end{subequations}
where
\begin{align}\label{ehmatrix}
\ve^{ij}&= \tilde{\varepsilon}^{ij}+\tilde{\gamma}^{i}_{\ k}\mu^{km}\tilde{\gamma}^{j}_{\ m}\, ,& \g^{ij}&=\tilde{\gamma}^{i}_{\ k}\mu^{kj}\, .
\end{align}

\section{Light rays and the Petrov direction}
\label{Lightrays}

We now analyze the trajectories of light rays using our decomposition. First we write \eqref{action} using \eqref{decomp},
\begin{equation}
 I=-\frac{1}{4} \int d^4 x \ \sqrt{-g} \  \Phi F^{\alpha\beta}F_{\alpha\beta}
-\frac{1}{8}\int d^4 x \ \sqrt{-g} \ W^{\alpha\beta\gamma\delta} F_{\alpha\beta}F_{\gamma\delta}.
\end{equation}
Varying with respect to the vector potential $A_\mu$ we obtain the Maxwell equations in the form,
\begin{equation}\label{max}
D_\alpha \left(\Phi F^{\alpha\beta} + \frac{1}{2}W^{\gamma\delta\alpha\beta}F_{\gamma\delta}\right)=0,
\end{equation}
where $D_\alpha$ stands for the covariant derivative with respect to the metric $g_{\mu\nu}$.
This type of modified Maxwell equations have been studied in other contexts. For instance,  to describe light propagation in black hole spacetimes backgrounds \cite{WeylM2,WeylM3}, and in holography \cite{WeylM4}, where the $W$-tensor is the Weyl tensor.

Here we will restrict our computations to the geometrical optics limit of Eq. \eqref{max}, valid when the wavelength $\l$ is much smaller than the typical length $L$ over which the properties of the wave and the medium change appreciably. We define a small parameter $\e = \l/L$, that will keep track of the corrections respect to the plane waves solutions.
In the the leading order of approximation we just write
\begin{equation}\label{goa}
A_{\m}=Re \left\{a_{\m}e^{iS/\e} \right\}
\end{equation}
where $a_{\m}$, $S$ are real, both of them functions of spacetime.
The wave vector $k^{\m}$ is defined by
\be{a}
k^{\mu}=\partial^{\mu}S.
\ee
We plug \eqref{goa} into Maxwell equation \eqref{max} and consider the leading term in this expansion.
\begin{equation}\label{O2m}
\Phi k^{\beta}k_{\beta}a^{\alpha}-k_{\beta}k_{\mu}W^{\alpha\beta\mu\nu}a_{\nu}=0.
\end{equation}
 If one sets $W^{\alpha\beta\mu\nu}=0$ and $\Phi=1$, one recovers the equations of a light ray in a gravitational field $g_{\m\n}$ (see \cite{gravitation}). We may also fix the gauge by demanding, for instance, the Lorenz gauge condition ${A^{\m}}_{;\m}=0$. To leading order in $\epsilon$  this gives
 \be{lorenz}
k_\m a^{\m}=0.
\ee

Now we concentrate in homogeneous materials, so that $W^{\alpha\beta\mu\nu}$ and $\Phi$ are constants. We will also take $g_{\mu\nu}$ to be the Minkowski metric $\eta_{\mu\nu}$ with mostly positive signature. We may use the methods of transformation optics \cite{chenchan,chen} to  later generate a broader family of materials.

It is well known that a tensor such as $W^{\alpha\beta\mu\nu}$ may be described using the  Petrov Classification \cite{Petrov:2000bs}. There are five non-trivial classes of such tensors.  Class I is the most generic form. The other four (named II, III, N and D) are called algebraically special, and have the property that there exist at least one vector $k_P^\mu$ such that
\be{special}
W_{\alpha\beta\mu\nu}k_P^\beta k_P^\nu = Ak_{P \a }k_{P \mu},
\ee
for some real number $A$. By virtue of \eqref{lorenz} and \eqref{O2m} we conclude that if a light ray has its wave vector  equal to $k^\mu_P$ then  $k_P^2\equiv \eta_{\mu\nu}k_P^\mu k_P^\nu=0$, that is, light rays traveling on that particular direction behave {\it as they were in Minkowski spacetime, at the speed of light in vacuum}. In this context, the axis along this direction is dubbed ``Minkowski axis''.

A key point to stress is that $k_P^{\mu}$ defines a direction on spacetime, therefore a ray moving in the opposite direction in space will not necessarily fulfill Eq. \eqref{special} (As we shall se bellow, this may only happen in the type D materials, where there are two linearly independent vectors $k_P^\mu$ satisfying \eqref{special}). These special directions underlines the key difference between a time-independent impedance matching media and one with non-zero $W$-tensor. The first must be  time symmetrical. Light rays traveling in one direction along a null geodesic should be able to travel in the opposite direction at the same speed. For the general case, however, this is not true anymore in general, because the action \eqref{action} is not necessarily time symmetrical even if the constitutive tensor is time independent. An interesting feature of the decomposition presented in this article is that it isolates  the part of the constitutive tensor which is responsible for asymmetrical propagation. Hence,  the study of this should  be useful in the design of this kind of materials.

We restrict in what follows to homogeneous materials belonging to an algebraically special Petrov class, and consider a cable extending along a Minkowski axis, $k_P^\mu$. This cable transmits light as vacuum does in one direction. In the other direction, in general, it will not, and one may take advantage of this feature. In the next section we study the properties of light  traveling in the reverse direction. It will be enough to consider the situation for a generic material belonging to each of the four non-trivial  algebraically special Petrov classes.

\section{The Petrov materials}
\label{PetrovNmaterialSe}
It is useful to work in a Newman-Penrose (NP) basis of null complex vectors $\{n^\mu,l^\mu,m^\mu,\bar m^\nu\}$, because there is a convenient  way of writing $W^{\a\b\g\d}$ in terms of this basis for each Petrov Class.  Here  the first two are real, taken so  that $n_\mu l^\mu=-1$.  The last two are complex and satisfy $m_\mu \bar m^\mu=1$, $m_\mu l^\mu=\bar m_\mu l^\mu=0$ and $m_\mu n^\mu=\bar m_\mu n^\mu=0$.
\subsection{Petrov N material}
\label{PetrovNmaterial}

A $W$-tensor of type N is written in a NP basis in the following form (see, for instance, \cite{stephani2004relativity}),
\be{N}
W^{\a\b\g\d}= \re \left[ 4\psi V^{\a\b} V^{\g\d} \right] =  2\left(\psi V^{\a\b} V^{\g\d} +  \bar\psi \bar V^{\a\b} \bar V^{\g\d}\right),
\ee
where $\psi =\alpha + i\beta$, is an arbitrary complex number ($\alpha,\beta$ real).
 The tensor $W$ may also be complex to describe energy gain or loss, as  we will discuss later. The tensor $V^{\alpha\beta}$ is given by
$$
V^{\a\b} = n^\a m^\b -n^\b m^\a.
$$
Of course $V^{\a\b}n_\b=0$,  showing that in this case $k_P^\mu = \sqrt{2}k n^\mu$  is the unique solution of \eqref{special}, for some wave number $k$.  We align $k_P^\mu$ with the $z$ axis, and choose the following NP-basis,
\be{NP}
n^\mu=\frac{1}{\sqrt{2}}\left(\begin{array}{c}
1 \\ 0 \\ 0 \\ 1
\end{array}\right) , \ \ \
l^\mu=\frac{1}{\sqrt{2}}\left(\begin{array}{c}
1 \\ 0 \\ 0 \\ -1	
\end{array}\right) , \ \ \
m^\mu=\frac{1}{\sqrt{2}}\left(\begin{array}{c}
0 \\ 1 \\ i \\ 0
\end{array}\right) , \ \ \
\bar m^\mu=\frac{1}{\sqrt{2}}\left(\begin{array}{c}
0 \\1 \\ -i \\ 0
\end{array}\right),
\ee
where we are using $$x^{\n}=\left(\begin{array}{c}t\\x\\y\\z \end{array}\right)\, .$$
The only non zero components of $W^{\a\b\g\d}$ turn out to be
\begin{eqnarray*}
W^{1313}&=&W^{1310}=W^{0101}=W^{2332}=W^{2302}=W^{2002}=\a,\\
W^{1323}&=&W^{1320}=W^{1023}=W^{0102}=\b.
\end{eqnarray*}
Actually, without loss of generality we can set $\beta=0$. This is because the basis vector $m^\alpha$ may be multiplied by the unit complex $e^{i\tau}$, for some real $\tau$ keeping the properties of the NP-basis. This is equivalent to perform a rotation of the NP tetrad around the $z$ axis. Thus, in Eq. \eqref{N} we are free to choose $m^\alpha$ in such a way that $\psi$ is real or, what is the same, to keep the basis as in \eqref{NP} rotating the material so that $\beta=0$.

The electric permitivity $\epsilon^{ij}$, the magnetic permeability $\mu^{ij}$ and the magneto-electric coupling $\gamma^i_j$, determining the chirality of the material, are computed from \eqref{ebmatrix},
\be{NPmatrixeh}
\ve=\left(\begin{array}{ccc}
\frac{1}{1+\alpha} &0&0 \\ 0&\frac{1}{1-\alpha}&0 \\ 0&0&1
\end{array}\right) , \quad \ \
\mu=\left(\begin{array}{ccc}
\frac{1}{1-\alpha} &0&0 \\ 0&\frac{1}{1+\alpha}&0 \\ 0&0&1
\end{array}\right) , \quad \ \
\g=\left(\begin{array}{ccc}
0&\frac{\alpha}{1+\alpha}&0 \\ \frac{\alpha}{1-\alpha}&0&0 \\ 0&0&0	
\end{array}\right) \, . \ \ \
\ee

We are now interested in what happens with light going in the backward direction. We define the wave vector
\be{back}
b^\mu = k\left(\begin{array}{c}
u \\ 0 \\ 0 \\ 1
\end{array}\right) = \frac{k(u+1)}{\sqrt{2}} n^\mu + \frac{k(u-1)}{\sqrt{2}}l^\mu,
\ee
where we take $k>0$.
Now we insert this in \eqref{O2m},
\begin{eqnarray}
0 &=& \Phi b^{\beta}b_{\beta}a^{\alpha}-b_{\beta}b_{\mu}W^{\alpha\beta\mu\nu}a_{\nu} \nonumber\\
&=& k^2 \left(\Phi (u^2-1)a^\alpha - 2(u-1)^2\alpha\re( m^\alpha m^\nu)a_\nu \right) \label{aa}.
\end{eqnarray}
Expanding $a^\mu$ in the NP-basis, we see from this equation that $a^\mu$ can have components along $m^\alpha$ and $\bar m^\alpha$  only. As $a^\mu$  must be real,
\be{AvectorNbasis}
a^\mu = \frac{1}{\sqrt{2}}\left(e^{i\theta} m^\mu + e^{-i\theta} \bar m^\mu\right),
\ee
which we choose to be unitary. Eq. \eqref{aa} has the solution $u=1$ for any polarization $a^{\mu}$. That was to be expected, because is the solution along $k_P^\mu$ we already discussed. If $u\neq 1$  \eqref{aa} yields
\be{equ}
\frac{u+1}{u-1}=\frac{\alpha}{\Phi}e^{-2i\theta},
\ee
and its corresponding complex conjugate. The fact that the phase velocity $u$ must be real, restricts $\theta$ to be either $0$ or $\pi/2$. In each case we obtain,
\be{ax}
u_x=-\frac{1+\alpha/\Phi}{1-\alpha/\Phi}, \ \   a_x^{\mu}= \left(\begin{array}{c}
0 \\ 1 \\ 0 \\ 0
\end{array}\right) ,  \ \ \ \
u_y=-\frac{1-\alpha/\Phi}{1+\alpha/\Phi}, \ \   a_y^{\mu}= \left(\begin{array}{c}
0 \\ 0 \\ -1 \\ 0
\end{array}\right) .
\ee
The solutions $(u_x, a^\mu_x)$, $(u_y, a^\mu_y)$ describe light polarized in the $x$ and $y$ axes respectively.

As long as $|\alpha/\Phi|<1$ these two phase velocities are  negative, representing light traveling in the $-\hat z$ direction. We note that the material is birefringent in that direction, with one polarization traveling faster than light in vacuum.

If $|\alpha/\Phi|>1$ then we obtain  phase velocities in the $\hat z$ direction, with one propagation being superluminal. This is not problematic, because it is the regime in which the media behaves as a metamaterial. This means that the Poynting vector points in the opposite direction to the phase velocity.
To see this, compute the energy-momentum tensor of the electromagnetic field,
\be{tmn}
T^{\mu\nu}= \bar H^{\mu\alpha}F^\nu_{\ \alpha} -\frac{1}{4}\eta^{\mu\nu}\bar H^{\alpha\beta}F_{\alpha\beta}.
\ee
The energy density of the electromagnetic field is
\be{T00}
\rho= T_{00}=-k^2 u,
\ee
while the Poynting vector $S^i=T^{0i}$ is
\be{T0i}
 {\bm S} = -u^2 k^2 \hat z,
\ee
showing that for $u>0$ the energy density is negative,
pointing always in the $-\hat z$ direction, independent of the direction of propagation defined by $u$.

\subsection{Petrov D material}

For the type D, the general $W$-tensor has the form \cite{stephani2004relativity}
\begin{equation}\label{PetrovD}
W^{\a \b \g \d}= Re \left[ 2\psi_D( V^{\a \b} U^{\g \d} + U^{ \a \b}V^{\g \d}+M^{\a \b}M^{\g \d})\right] ,
\end{equation}
where $\psi_D=\alpha_D+i\beta_D$ and, 
\be{UM}U^{\a \b}=-l^\a \bar m^\b+l^\b \bar m^\a, \ \   \ \ \ M^{\a \b}=m^{\a}\bar m^{\b}-m^{\b}\bar m^{\a}-n^{\a}l^{\b}+n^{\b}l^{\a}.
\ee
 In this case, both $k_P^\mu= \sqrt{2}k n^\mu$ or $k_P^\mu=\sqrt{2}k l^\mu$ satisfy \eqref{special}, and therefore light may travel along the $z$ axis in both directions, and with any polarization at the speed of light $u=\pm 1$.

For this case, the matrices  \eqref{ebmatrix}   are
\be{NPmatrixeh}
\ve=\left(\begin{array}{ccc}
\frac{(1+\alpha_D)^2+\beta^2_D}{1+\alpha_D} &0&0 \\ 0& \frac{(1+\alpha_D)^2+\beta^2_D}{1+\alpha_D}& 0 \\ 0&0&\frac{(1-2\alpha_D)^2+4\beta^2_D}{1-2\alpha_D}
\end{array}\right) , \ \
\mu=\left(\begin{array}{ccc}
\frac{1}{1+\alpha_D} &0 &0 \\ 0 &\frac{1}{1+\alpha_D}&0 \\ 0&0&\frac{1}{1-2\alpha_D}
\end{array}\right) , \quad \ \
\ee
\be{gD}
\g=\beta_D \left(\begin{array}{ccc}
-\frac{1}{1+\alpha_D}&0&0 \\ 0&-\frac{1}{1+\alpha_D}&0 \\ 0&0&\frac{2}{1-2\alpha_D}
\end{array}\right) \, . \ \ \
\ee

\subsection{Petrov III material}

For the type III materials, we find
\begin{equation}\label{Tres}
W^{\a \b \g \d}= Re \left[ 4\psi V^{\a \b} V^{\g \d} + 2\psi_3(V^{ \a \b}M^{\g \d}+M^{\a \b}V^{\g \d})\right],
\end{equation}
where $\psi=\alpha$ is again real, and $\psi_3=\alpha_3+i \beta_3$. Here $k_P^\mu= \sqrt{2}k n^\mu$  satisfy \eqref{special}, hence light travels in the direction of $+\hat z$ with any polarization at the speed of light. In the direction $-\hat z$, however, the material is birefringent, and the different speeds of light are
$$
u_x=\frac{1+\alpha^2+4\alpha(\alpha_3^2-\beta_3^2)+2\sqrt{\left(\alpha + 2(\alpha_3^2 - \beta_3^2 ) \right)^2 + 16\alpha_3^2\beta_3^2}}{(\alpha^2-1+4\alpha(\alpha_3^2-\beta_3^2)+4(\alpha_3^2+\beta_3^2))},
$$
$$
u_y=\frac{1+\alpha^2+4\alpha(\alpha_3^2-\beta_3^2)-2\sqrt{\left(\alpha + 2(\alpha_3^2 - \beta_3^2 ) \right)^2 + 16\alpha_3^2\beta_3^2}}{(\alpha^2-1+4\alpha(\alpha_3^2-\beta_3^2)+4(\alpha_3^2+\beta_3^2))}
$$
for $x$-polarized and $y$-polarized light respectively.
In general these expressions are not quite enlightening. Note however, that in the limit of $\alpha\gg\alpha_3,\beta_3$ we recover the expressions \eqref{ax} for the type N material. It is interesting, therefore, to explore the opposite limit, $\alpha\rightarrow 0$ , in which the identity of the type III material is better captured.
In that case we obtain,
\be{uIII}
u_x = -\frac{1+\sigma^2}{1-\sigma^2}, \ \ \ \ u_y = -1,
\ee
where $\sigma^2 = 2(\alpha_3^2+\beta_3^2)$. This case is quite interesting. In the $-\hat z$ direction the material is birefringent, but for one $y$-polarization, light travels at the speed of light. If only this polarization is used, the material behaves exactly like the type D material:  light travels as in Minkowski both ways. If $x$-polarized light is used instead, the behavior resembles the one for the N type material: light travels as in Minkowski one way, but not in the opposite, where its speed is always greater than light in vacuum. When $\sigma>1$, it describes a metamaterial. 

In the present case, we display the matrices  \eqref{ebmatrix} in the just described case of $\alpha=0$, taking also, for simplicity, the case of $\beta_3=0$,
\be{matricesII}
\ve=\left(\begin{array}{ccc}
1&0&-\alpha_3 \\ 0& \frac{1}{1-\alpha_3^2}& 0 \\ -\alpha_3&0&1+\alpha_3^2
\end{array}\right) , \ \ \ \ \ \ \ \ \ \ 
\mu=\left(\begin{array}{ccc}
\frac{1}{1-\alpha_3^2} &0 &\frac{\alpha_3}{1-\alpha_3^2} \\ 0 & 1 &0 \\ \frac{\alpha_3}{1-\alpha_3^2}&0&\frac{1}{1-\alpha_3^2}
\end{array}\right) , \quad \ \
\ee
\be{gD}
\g= \left(\begin{array}{ccc}
0&0&0 \\ \frac{\alpha_3^2}{1-\alpha_3^2}&0&\frac{\alpha_3}{1-\alpha_3^2} \\ 0&\alpha_3&0
\end{array}\right) \, . \ \ \
\ee

\subsection{Petrov II material}

Finally, a type II material may be described by adding up the $W^{\a \b \g \d}$ tensors for type D in \eqref{PetrovD} and for type III in \eqref{Tres}.
 Again, $k_P^\mu= \sqrt{2}k n^\mu$  is the unique vector satisfying \eqref{special}, therefore  light directed along $+\hat z$ with any polarization travels at the speed of light in vacuum.
For the light traveling in the inverse direction the situation is much more involved, because there are five different real parameters to adjust. Two in $\psi_D$ in \eqref{PetrovD}, and four in $\psi$ and $\psi_3$ in \eqref{Tres}. We may, however, take one of them, say $\psi=\alpha$ to be real. Of course, by choosing the appropriate parameters very close to zero, we may approach to types N, D and III as close as we want.
Therefore  we will only  present a particular case, which best illustrates the remarkable features of this type of material. We take $\psi_3=i\beta_3$, $\psi_D=i\beta_D$ and $\psi=0$. It is straightforward to show that for one polarization light moves with the speed of light in vacuum, while for its orthogonal counterpart the phase velocity is given by
$$
u=\frac {\beta_3^{2}-\beta_D^2+1}{\beta_3^{2}+\beta_D^2-1}.
$$
When $\beta_D^2>1$ one obtains that $|u|<1$. This is complementary to the case III above. For one polarization light travels as in Minkowski both ways. For the other, the behavior resembles the one for the N type material: light travels as in Minkowski one way, but not in the opposite, where its speed is always {\it smaller} than the speed of light in vacuum. When $\sigma>1$ it describes a metamaterial.

In the present case the matrices  \eqref{ebmatrix} are
\begin{equation*}
\ve=\left(\begin{array}{ccc}
\frac{1+\b_D^2-\b_3^2\b_D^2}{1-\b_3^2}&\frac{\b_3^2\b_D}{1-\b_3^2}&\frac{(1+\b_3^2)\b_3\b_D}{1-\b_3^2}\\\frac{\b_3^2\b_D}{1-\b_3^2}&\frac{1-\b_3^2+\b_D^2}{1-\b_3^2}&\frac{\b_3(1-\b_3^2+\b_D^2)}{1-\b_3^2}\\\frac{(1+\b_3^2)\b_3\b_D}{1-\b_3^2}&\frac{\b_3(1-\b_3^2+\b_D^2)}{1-\b_3^2}&\frac{1-\b_3^2(\b_3^2-1)}{1-\b_3^2}
\end{array}\right) ,
\end{equation*}
\begin{equation*}
\mu=\left(\begin{array}{ccc}
1&0&0\\0&\frac{1}{1-\b_3^2}&\frac{-\b_3}{1-\b_3^2}\\0&\frac{-\b_3}{1-\b_3^2}&\frac{1}{1-\b_3^2}
\end{array}\right) ,
\end{equation*}
\begin{equation*}
\g=\left(\begin{array}{ccc}
-\b_D&-\frac{\b_3^2}{1-\b_3^2}&-\frac{\b_3}{1-\b_3^2}\\0&-\frac{\b_D}{1-\b_3^2}&\frac{\b_3\b_D}{1-\b_3^2}\\ \b_3&-\frac{\b_3\b_D}{1-\b_3^2}&\frac{\b_D}{1-\b_3^2}
\end{array}\right) .
\end{equation*}

\subsection{Loss and gain}

Until now, we have only discussed the case of a lossless media. To describe a more general situation one has to  relax the requirement that the $W$-tensor is real. We will proceed in  the case of Petrov N material of Sec. \ref{PetrovNmaterial} for simplicity. Other cases can be studied straightforwardly. We write the complex tensor
\be{N2}
W^{\a\b\g\d}=2\left( \psi_1 V^{\a\b} V^{\g\d} + \psi_2 \bar V^{\a\b} \bar V^{\g\d}\right),
\ee
as a generalization of the tensor \eqref{N}, where $\psi_1 =\alpha_1$ and $\psi_2 =\alpha_2 + i\beta_2$,  with $\alpha_1, \alpha_2, \beta_2$ all real. Here $V^{\a\b}$ is the same antisymmetric tensor defined previously. Notice that it is not possible to eliminate the imaginary part of $\psi_2$ redefining the vector $m^\alpha$
by some factor $e^{i\tau}$.

For the $W$-tensor \eqref{N2}, any wave vector $k^\alpha\propto n^\alpha$ solves Eq. \eqref{special} defining a ray moving at the speed of light. We are again interested in the backward propagation of light and if there are some difference produced by the complex nature of the $W$-tensor.
We can use the same wave vector \eqref{back}, but requiring that the frequency $\omega$ be a complex scalar. Repeating the same analysis performed in Sec. \ref{PetrovNmaterial}, Eq. \eqref{O2m} yields
\be{aa22}
0= \Phi (\omega^2-k^2)a^\alpha - (\omega-k)^2\psi_1 m^\alpha m^\nu a_\nu - (\omega-k)^2\psi_2 \bar m^\alpha \bar m^\nu a_\nu,
\ee
which allows us to determine  $a^\mu$  as the real unitary vector \eqref{AvectorNbasis} in the NP-basis.
With this, Eq. \eqref{aa22} produces the following equations
\begin{eqnarray}
\omega + k &=& \frac{1}{\Phi} \left(\alpha_2+i\beta_2\right) (\omega-k)e^{2i\theta},\\
\omega + k &=& \frac{\alpha_1}{\Phi}(\omega-k)e^{-2i\theta}.
\end{eqnarray}
These two equations can be solved as
\be{omegakcomplejo}
\omega+k=\pm(z_1+iz_2)(\omega-k),
\ee
where $z_1+iz_2=\sqrt{\alpha_1(\alpha_2+i\beta_2)}/\Phi$.
In the limit $\beta_2=0$,  and $\alpha_1=\alpha_2$ (implying $z_2=0$ and real frequencies),  we recover the previous results of Sec.~\ref{PetrovNmaterial}.

The frequency $\omega$ is now a complex quantity, which can be written as $\omega=\omega_R+i\eta$. Its imaginary part $\eta$ produces damping or instability of the wave,  as it couples to the amplitude vector in the form $e^{-\eta t}$. Depending on the sign of $\eta$, the amplitude will decrease or increase on time. There are two solutions of Eq. \eqref{omegakcomplejo}, depending on the wave polarization. These are
\be{}
u_{\pm}=\frac{\omega_R}{k}=-\frac{1-{z_1}^2-{z_2}^2}{\left(1\mp z_1\right)^2+{z_2}^2},\qquad
\eta_{\pm}=\frac{\mp 2 k z_2}{\left(1\mp z_1\right)^2+{z_2}^2}.
\ee

We can see that the complex nature of the $W$-tensor modifies the propagation of light for each polarization. Interestingly enough, the electromagnetic wave can be damped or it can grow due to the imaginary parts of the $W$-tensor. We can study  a small departure with respect to the real case of Sec.~\ref{PetrovNmaterial}. If the imaginary part is small enough $z_1\gg z_2$,  then
\be{}
u_{\pm}\approx-\frac{1-{z_1}^2}{\left(1\mp z_1\right)^2}\, ,\qquad
\eta_{\pm}\approx\frac{\mp 2 k z_2}{\left(1\mp z_1\right)^2},
\ee
where the phase velocities for both polarizations coincide with those of Sec.~\ref{PetrovNmaterial}. In the case $z_1$ approaches to 1, the magnitude of the phase velocity can become extremely large. However, the imaginary part can act to forbid that behavior. When $z_1\approx 1-\epsilon$, with $1\gg\epsilon$, then $-\sqrt{2\epsilon}<z_2<\sqrt{2\epsilon}$, to preserve the negative phase velocity.  This implies that
\begin{eqnarray}
u_{+}&\approx&-\frac{2}{\epsilon}\, ,\qquad \frac{2\sqrt{2}}{\epsilon^{3/2}} > \frac{\eta_{+}}{k}> - \frac{2\sqrt{2}}{\epsilon^{3/2}}\, ,\nonumber\\
u_{-}&\approx&-\frac{\epsilon}{2}\, ,\qquad \sqrt{\frac{\epsilon}{2}} > \frac{\eta_{-}}{k}> -\sqrt{\frac{\epsilon}{2}}\, .
\end{eqnarray}
Therefore, if $-\sqrt{2\epsilon}<z_2<0$, for one polarization the wave is damped as its phase velocity increases in the backward direction. For the other polarization the wave amplitude increases as it slows down, but this increasing is insignificant
as the wave travels slower.

\section{Conclusions}
\label{discusion}

The introduction of an appropriate decomposition of the constitutive tensor through the $W$-tensor \eqref{decomp} generalizes the expression \eqref{metricdecomp} which is valid for impedance matching materials, that may be treated using transformation optics.  The new term, dubbed the $W$-tensor, may be used to analyze, using the Petrov classification, the possible asymmetric transmission exhibited by a homogenous cable made out of this material. In particular, we study the case in which light travels as in vacuum along a preferred direction, whereas,  in the opposite direction, it may travel slower, faster or at the same speed that light in vacuum, depending on its Petrov type and its polarization. The fact that all the quantities introduced in the new decomposition are tensorial, make it possible to use it in conjunction with transformation optics (we know how every quantity transforms under a coordinate transformation). The methods presented here may be useful in the design of a new kind of devices, such as light diodes.

\section*{Funding}

The Centro de Estudios Cient\'ificos (CECs) is funded by the Chilean Government
through the Centers of Excellence Base Financing Program of Conicyt. The work of C.E. was supported by Conicyt. The work of F.A.A. was partially supported by Fondecyt--Chile Grant No. 11140025. The work of A.G. was partially supported by Fondecyt--Chile Grant No. 1141309.

\end{document}